\documentclass[preprint,journal]{vgtc}       

\ifpdf
  \pdfoutput=1\relax                   
  \usepackage{graphicx}                
  \DeclareGraphicsExtensions{.pdf,.png,.jpg,.jpeg} 
\else
  \usepackage{graphicx}                
  \DeclareGraphicsExtensions{.eps}     
\fi%

\graphicspath{{figures/}{pictures/}{images/}{./}} 

\usepackage{microtype}                 
\PassOptionsToPackage{warn}{textcomp}  
\usepackage{textcomp}                  
\usepackage{mathptmx}                  
\usepackage{times}                     
\usepackage{cite}                      
\usepackage{tabu}                      
\usepackage{booktabs}                  

\usepackage{graphicx}
\usepackage{tabularx}
\usepackage{soul,color}
\usepackage{soul}
\usepackage{float}
\usepackage{adjustbox}
\usepackage{enumitem}

\preprinttext{Accepted at IEEE Vis 2022. To appear in IEEE Transactions on Visualization and Computer Graphics.}

\onlineid{1161}

\title{PMU Tracker: A Visualization Platform for Epicentric\\Event Propagation Analysis in the Power Grid}

\author{Anjana Arunkumar, \textit{Student Member, IEEE}, Andrea Pinceti, \\Lalitha Sankar, \textit{Member, IEEE}, and Chris Bryan, \textit{Member, IEEE}}
\authorfooter{
The authors are with Arizona State University: \{ aarunku5, apinceti, lsankar, cbryan16 \}@asu.edu
}

\newcommand{\name}{PMU Tracker}

\abstract{The electrical power grid is a critical infrastructure, with disruptions in transmission having severe repercussions on daily activities, across multiple sectors. To identify, prevent, and mitigate such events, power grids are being refurbished as `smart' systems that include the widespread deployment of GPS-enabled phasor measurement units (PMUs). PMUs provide fast, precise, and time-synchronized measurements of voltage and current, enabling real-time wide-area monitoring and control. However, the potential benefits of PMUs, for analyzing grid events like abnormal power oscillations and load fluctuations, are hindered by the fact that these sensors produce large, concurrent volumes of noisy data. In this paper, we describe working with power grid engineers to investigate how this problem can be addressed from a visual analytics perspective. As a result, we have developed PMU Tracker, an event localization tool that supports power grid operators in visually analyzing and identifying power grid events and tracking their propagation through the power grid's network. As a part of the PMU Tracker interface, we develop a novel visualization technique which we term an \textit{epicentric cluster dendrogram}, which allows operators to analyze the effects of an event as it propagates outwards from a source location. We robustly validate PMU Tracker with: (1) a usage scenario demonstrating how PMU Tracker can be used to analyze anomalous grid events, and (2) case studies with power grid operators using a real-world interconnection dataset. Our results indicate that PMU Tracker effectively supports the analysis of power grid events; we also demonstrate and discuss how PMU Tracker's visual analytics approach can be generalized to other domains composed of time-varying networks with epicentric event characteristics.
}

\keywords{Human-centered computing, Dendrograms, Visualization design and evaluation methods, Cyber-physical networks}

\CCScatlist{
  \CCScatTwelve{Human-centered computing}{Visu\-al\-iza\-tion}{Visu\-al\-iza\-tion techniques}{Dendrograms};
  \CCScatTwelve{Human-centered computing}{Visu\-al\-iza\-tion}{Visualization design and evaluation methods}{};
  \CCScatTwelve{Networks}{Network Types}{Cyber-physical networks}{Sensor networks};
}

\renewcommand{\manuscriptnotetxt}{}

\vgtcinsertpkg

\begin{document}
\firstsection{Introduction}
\maketitle

As a critical infrastructure component, modern electric power grids are large, distributed, and complex cyber-physical systems employing real-time wide-area monitoring and control to ensure reliable energy delivery.  In recent years, grids are experiencing heightened complexity and unpredictability, due to congestion, atypical power flows, and increasing demand for renewable energy~\cite{1507024}. To enhance resilience and support real-time decision-making, power systems have accordingly been revamped as \textit{smart grids}\cite{6099519}, with the decentralized, autonomous operation of bus controllers to develop self-healing and interoperable grids~\cite{tebekaemi2019secure} that integrate varied energy sources~\cite{worighi2019integrating}. 

One facet of grid modernization is the increasing deployment of GPS-enabled phasor measurement units (PMUs)~\cite{eia} by industry stakeholders, which provide high-speed, high-resolution, synchronized measurements of direct voltage and current phasor measurements~\cite{mauryan2014phasor}. While PMUs allow for precise assessment of power flows and power quality~\cite{6036498}, their adoption and usage by power system operators is hindered due to several factors including cost, cybersecurity, human operators, etc. In the context of analyzing grid events like abnormal power oscillations, line losses, and load fluctuations, the limited usage of PMU data can be attributed (in part) due to high (noisy) data volumes and high sampling rates~\cite{8245272}.

Visual analysis of grid events generally combines backend processes with frontend techniques that show identified anomalies. Unfortunately, current research and industry platforms (e.g.,~\cite{6698979, 7856726,cozby2019visualization}) afford limited scalability and flexibility for event analysis. Indeed, for many operators, tracking the occurrence and progression of grid events involves a tedious manual mapping process, whereby line charts showing raw data streams (at the rate of 30-120Hz) must be cross-referenced with network diagrams of the power grid components to understand how anomalous behavior is progressing through the network's topology. This constrains operator ability to conduct real-time decision-making and hinders their ability to flag secondary events further downstream of the event source where they can cause significantly more disruption (such as blackouts).

As a solution, in this paper, we implement \textit{PMU Tracker}, an end-to-end visual analysis and event localization tool that affords streamlined, flexible, and scalable analysis of PMU data. The high-level workflow supported by the system is as follows: based on an encountered (or identified) grid event, users can apply spectral analysis for anomaly detection across a set of coordinated visualizations. In particular, the ``event epicenter'' (the PMU(s) at or near the center of the event) can be identified and highlighted for further analysis, and nearby PMUs can be selected, clustered, and tracked to understand how the event propagates out from the source into the surrounding network. This is enabled by our development of a novel \textit{epicentric cluster dendrogram} visualization, which affords users the ability to interactively track event propagation from the epicentric PMU(s).

In this paper, we describe the process of developing PMU Tracker, motivated by working with grid engineers and conduct a duo of evaluations (a usage scenario, and case studies with domain experts) to validate both the overall tool and the novel epicentric cluster dendrogram technique's ability to support analysis of grid events. 
Moreover, since PMUs function as sensors in the cyber-physical power grid system, we also briefly discuss the generalizability of PMU Tracker to other time-varying sensor networks that contain epicentric or localized events.

Succinctly, this paper's contributions can be summarized as follows:
(1)~Based on surveying a set of experienced practitioners who work for U.S. power companies to understand industry challenges and practices for handling and visualizing PMU data, we identify a set of design requirements for effectively visualizing and analyzing how events propagate through the grid network.
(2)~Based on these requirements, we implement a software platform called PMU Tracker to visually identify and analyze events, and to topologically track event propagation through the power grid network. PMU Tracker includes both a data management backend (using the Parquet file format) and a coordinated interface consisting of several linked visualizations.
(3)~As a part of the PMU Tracker interface, we develop a novel epicentric cluster dendrogram view to spatiotemporally analyze topological event propagation through the power grid network, by juxtaposing selected PMUs against the event's epicenter PMU(s). (4)~Based on evaluating PMU Tracker, we demonstrate and discuss how visual analytics can be used to analyze power grid event anomalies. Further, we demonstrate how PMU Tracker can be extended to additional, related sensor-based domains that have epicentric event considerations.
\section{Related Work}

This work sits at the intersection of two primary areas: (1) \textit{visual analysis of the electric power grid}, whereby domain operators and researchers visualize grid topology and data streams (including PMU data), and (2) \textit{time-varying topological visualization}, where we consider networks with attribute values that change over time, and the topology of the network is an important consideration for analysis.

\subsection{Visual Analysis of the Electric Power Grid}

PMUs are measurement devices used in power systems that provide complex measurements (phase and magnitude) of voltages and currents~\cite{monti2016phasor}. Measurements (usually 30--120 samples per second) are GPS-synchronized across wide geographical areas, providing a precise picture of the grid state. PMUs enable the capture of dynamic subsecond behaviors that were previously unobservable on the grid. The adoption of PMU-equipped substations by electric utilities, however, has been slow and scattered due to high communication and data processing infrastructure requirements, as well as large demands for investment and operator training. This necessitates new visual tools that can streamline the analysis of PMU data to purvey its practical application in event propagation tracking.

\textbf{Visualization Techniques for Streaming Power Grid Data.}
Two survey papers~\cite{nga2012visualization,sanchez2018survey} on domain visualization techniques for smart grids provide an overview of techniques, which are commonly used in industry, including line charts for temporal data and node-link diagrams for network topology. 

Several industry and open-source software platforms exist for capturing and visualizing PMU and SCADA data streams. Real-Time Dynamics Monitoring System~\cite{6863485}, or RTDMS, is a platform for real-time grid monitoring that includes many ``familiar'' views that are standard to domain operators: geographic displays overlaid with the grid topology to show network state and connectivity, line charts for raw PMU data streams, and odometer-inspired radial dials for phase angle separation, voltage sensitivity, oscillation, and grid stress monitoring. Other wide-area monitoring and control systems include Babazadeh et al.~\cite{6652202} and Cozby et al.~\cite{cozby2019visualization}. While such systems are adept at capturing and visualizing PMU data streams in real-time, and include backend algorithms to identify anomalous behavior such as oscillations or line faults, they generally do not provide visualization support specifically tailored for event analysis. In contrast, while PMU Tracker currently supports near-time analysis using a historical PMU dataset, it is specifically designed to support event analysis (by focusing on epicentric protocols used by operators when responding to system events) and its views can be adapted to work for real-time data.

\textbf{Encoding Power Dynamics: } Several platforms have extended the base ``geographical map with overlaid node-link diagram'' technique to encode additional information. For instance, a contour map can be overlaid to represent bus and line voltage/current profiles of regions either geographically or within bus systems~\cite{sheikholeslami2017visualization,7459263}. 
Transforming the network's node-link diagram to use 3D glyphs has also been employed in several platforms~\cite{885101,pienta2016steps}, as well as the use of animated line flow visualizations superimposed atop maps, and overlaying pie charts at line junctions to represent loading levels~\cite{926744}.

Unfortunately, such layered visualizations are generally either cluttered or utilize suboptimal design choices (e.g., many contour maps employ rainbow color scales). We note this not to say that existing systems are ineffective or wrong (to the contrary, many employ significant software engineering to enable analysis of large and streaming grid data), but to point out that (1)~PMU Tracker strives to employ best practices for visualization design choices, and (2)~future visualization-based tools for the power grid can likely be improved by emphasizing collaboration with the visualization and design communities.

\textbf{Visualizing Grid Events.}
Visual analysis of grid events (line/generation losses, voltage drops, oscillations, etc.) generally combines backend processes with frontend techniques that show identified anomalies. For instance, PMUs are flagged when monitored values exceed thresholds, network connectivity drops between components, or anomalies are detected using ad hoc algorithms~\cite{8353135}. Current methods require high operator knowledge and processing overhead; this regularly results in flagging of events further downstream than when anomalies first occur, where they can cause significantly more disruption (such as blackouts). This helps to motivate the current work: by developing epicentric visualization designs, we hope to streamline event analysis and support identification of downstream events.

Signal processing methods such as fast Fourier transform (FFT), matrix-pencil, and spectral analysis~\cite{1717580} are a common approach for characterizing the time evolution of PMUs that are potentially relevant to an oscillation event (i.e., abnormally high frequency and magnitude load fluctuations). For example, Messina et al.~\cite{1717580} used Hilbert spectral analysis to visualize (using line charts) and characterize nonlinear oscillations from synchronized wide-area measurements, as a way to help determine event propagation through the network. Similarly, PMU Tracker allows interactive signal processing of PMU data to analyze grid events, where selected PMU streams are decomposed using FFT. (Additional event analysis techniques can be integrated into the system as future work, see the Discussion section). In contrast to the above systems, we additionally provide epicentric visualization and analysis, which is aimed at explicating customized ranking methods~\cite{Mishra2021HowRA} (based on relevance to the event under consideration) for PMUs.

\subsection{Time-Varying Topological Visualization}

Spatiotemporal visualization concerns changes in information in space and time, and allows for the identification of overall trends and movement patterns. Space-time cubes \cite{langran1988framework} show how phenomena change over time within geographic space, and have been used for representing time-dependent map data \cite{calabrese2014urban}. Importance-driven visualization of time-varying data \cite{4658174} and its optimization for remote sensing \cite{fellegara2021interactive} have also been investigated. Other techniques include sequential snapshots\cite{beck2013matching}, base-state with amendments (such as changing contour over time), and space-time composites \cite{javed2012exploring}.

Anomaly detection in spatio-temporal networks for smart cities has been implemented in systems like Voila \cite{8022952} and EnsembleLens \cite{8440102} based on these considerations. Flow maps have also been proposed, where discrete spatiotemporal data is represented as a continuous function (KDE), and flow maps are extracted using a 3D gravity model for temporal trends (considered as movement flows) \cite{7847429}. Scalability-centric approaches (e.g., VIVA~\cite{6557147}) have suggested the use of timeline visualizations, in association with network data, to construct node glyphs for representing resource flow characteristics in large-scale distributed systems.

Notwithstanding the contribution of such approaches towards anomalous pattern identification in time-varying networks, their extensibility to the power domain is limited by several factors: (1) the presence of unknown or ambiguous potential event locations, (2) inability to identify concurrently occurring events, and (3) scalability and flexibility in juxtaposing network nodes against nodes of interest to track the evolution and propagation of anomalies. PMU Tracker addresses these issues via the development of a novel dendrogram cluster visualization to enable epicentric event propagation analysis.
\section{Domain Survey and Task Analysis}
\label{sec:domain_and_task_analysis}

To help motivate our software development efforts, we surveyed three power system engineers that we have previously collaborated with to understand current challenges for visualizing and analyzing PMU-based power grid data. These industry professionals have extensive experience working for U.S. power companies (13, 15, and 32 years) in contingency analysis, systems protection, and control and coordination.

Table~\ref{tab:survey_challenges} lists several challenges explicitly mentioned by the practitioners for visualizing PMU data. While all three practitioners regularly use a combination of general-purpose scripting tools/languages and in-house/dedicated visualization software (RTDMS, ASPEN, PSS/E, etc.) to analyze both historical and real-time grid data, they each described event analysis workflows as ``\textit{unintuitive}.'' Often, event analysis involved reviewing line charts of the PMU data and manually matching PMUs to their network/geographic locations. Event-related visualizations are generally manually created, such as via scripting languages (R and Python were commonly referenced). Such a manual workflow imposes a significant overhead on operators, limiting their ability to efficiently analyze anomalous data in the grid. 

\begin{table}[!h]
    \centering
    \footnotesize
    \begin{tabularx}{\columnwidth}{XlX}
    \toprule
    \multicolumn{2}{l}{\textbf{Challenges for visualizing PMU data}} & \\
    \midrule
         (3) & Quick and efficient data retrieval. & *  \\
         (2) & Visualizing multiple PMUs simultaneously and isolating outliers. & * \\
         (2) & Tracing event propagation through the network. & * \\
         (1) & Linking data to events and identifying missing data. & * \\
         (1) & Streaming PMU data for real-time decision-making. \\
    \midrule
    \end{tabularx}
    \caption{Our survey with power systems professionals identified several items as salient challenges for visualizing power grid data. We note the number of users who referenced each challenge in parentheses. Items addressed by our current tool are noted with an asterisk.}
    \label{tab:survey_challenges}
\end{table}

Based on the challenges listed in Table~\ref{tab:survey_challenges}, we defined a set of task requirements that a visual analytics system can employ for analyzing and tracking the evolution of localized grid events.

Such a system could potentially benefit power systems engineers including those engaged in (i) contingency analysis, (ii) systems protection, control, and coordination, (iii) line engineering based on historical event data, (iv) handling transient and dynamic stability in the grid, and (v) the analysis of disturbance events, misoperation events, and synchrophasor data.

\textbf{T1: Flexible and interactive data retrieval.} Systems should provide a backend that supports real-time, customizable data retrieval to support flexible front-end exploration and enable early event flagging.

\textbf{T2: Link events to PMUs.} Localized events can be sourced to a specific region of the power grid network, either via automatic means~\cite{8082533,8353135,4275462} or by manually identifying one or more epicenter PMUs, which are adjacent to or at the event source, to begin tracking event propagation throughout the network. 

\textbf{T3: Support temporal analysis.} As PMUs provide fine-grained data sampling (30 Hz or higher), subsecond events can be traced over the network, though sometimes events can have a duration effect of multiple hours. As a result, visualizations should flexibly support temporal analysis at multiple scales.

\textbf{T4: Provide multiple perspectives for analyzing events.} 
Providing multiple coordinated visualizations that support different data perspectives can provide a robust view for analysts who are identifying and analyzing an event. In particular, as many grid events begin as localized phenomena that spread throughout the network, including an epicentric view can potentially provide important analytic benefits to understand propagation. In our case, we incorporate views that support signal processing of temporal data streams, a global view of the grid network, a similarity-based plot of PMUs, and the novel dendrogram plot for epicentric analysis.

In contextualizing \textbf{T1}--\textbf{T4} to existing grid software, to our knowledge no existing industry or research platforms wholly address all of these tasks. In this way, our system represents a scalable, end-to-end spatiotemporal solution for event localization and propagation tracing over a power grid.

\section{PMU Dataset Storage and Retrieval}
\label{sec:data_management}

PMU data is large-scale, noisy, and multi-attribute, but to support task \textbf{T1}, the retrieval of PMU data should be fast, flexible, and interactive. In this paper, we use a dataset that comprises $\sim$500 PMUs over a three-year period from a regional U.S. gas and electric company.\footnote{Access to this PMU dataset is provided under NDA, so we anonymize certain features in this paper, such as names, IDs, and locations, and only reference the approximate number of PMUs available.} 
This dataset additionally contains a collection of operator reports about historical grid events, stored as unstructured text files.

We save the raw PMU data to disk on a compute cluster using the Parquet file format. Parquet is a columnar storage scheme based on an algorithm for record shredding and assembly~\cite{36632}. It supports efficient and flexible compression and encoding. A Parquet file is stored as a series of blocks (row-groups), each of which in turn contains column groups; the values of the column groups are stored in contiguous memory locations. This transparent column-specific compression allows for high-performance data fetching compared to row-wise queries.

\begin{figure}[!h]
    \centering
    \includegraphics[width=\columnwidth]{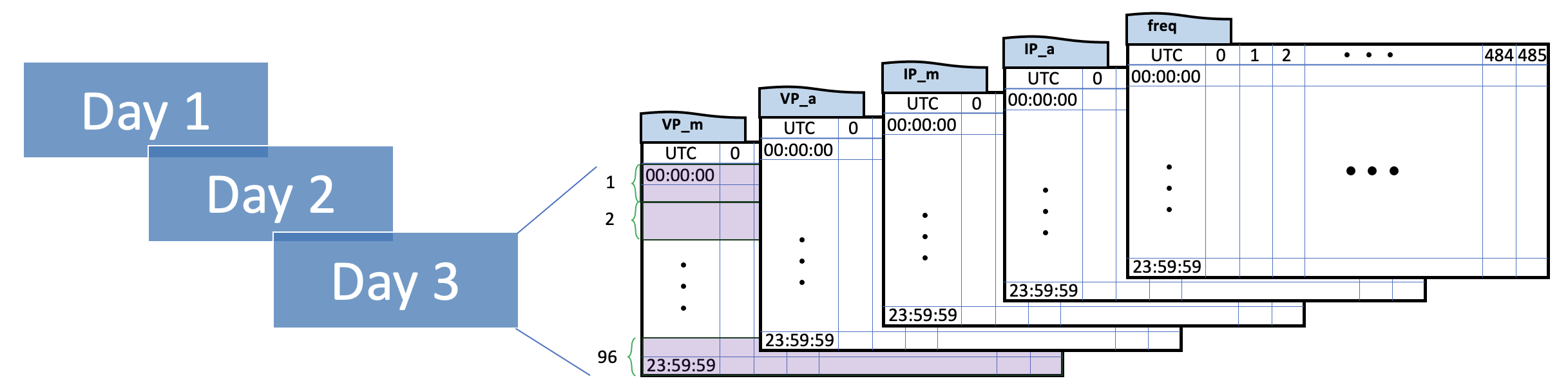}
    \caption{The storage format for PMU data. Raw data is saved into Parquet format (1 file equals 1 PMU for 1 day), allowing fast column-wise reads for small time durations. Attributes stored are the magnitude and phase data of voltages and currents, at 30Hz frequency.}
    \label{fig:parquet_setup}
\end{figure}

In our dataset, each Parquet file stores one attribute's worth of data for all PMUs for one day (using 96 fifteen-minute groups), see Figure~\ref{fig:parquet_setup}. 
The PMUs in our dataset record measurements at 30 Hz (during data dropout, null values are recorded), meaning each Parquet file is $\sim$500 columns $\times$ 2.592M rows. As the PMUs in our dataset record 18 attributes (with raw values stored as 64-bit floating-point values), each day contains 18 PMU files. 
In this manner, we optimize our storage for fast querying of PMUs for a single attribute of interest; for example, loading all $\sim$500 PMUs for a single day for one attribute requires just under six seconds (for one PMU: under two seconds). Query times scale approximately linearly as additional attributes are retrieved.

PMU data stored in Parquet files is compressed and decompressed automatically and on-the-fly using Gzip. The compressed size of our Parquet files can vary between attributes. For example, in  compressed format, files for the \textsf{Magnitude} attribute are generally between 1.5--2.5 GB. When reading data from a Parquet file, uncompressing a row group results in reading approximately 20 MB of data. Storage-wise, each day's set of 18 Parquet files is approximately 50--70 GB, with the full PMU dataset for three years being over 70 TB. There are current efforts to add an additional six years of PMU data to this dataset, which will equal nine years of data totaling over 200 TB in Parquet format.

Our PMU dataset additionally contains a collection of reports created by operators that describe grid events of interest, such as oscillations or faults. Reports are stored as unstructured text documents, meaning they do not provide any metadata explicitly linking the report to PMUs, substations, or locations in the grid (report timestamps are the only metadata). To link specific events to one or more epicenter PMUs, we use a regex matching schema between report text and PMU IDs, substation names (where the PMU is located), and their locations. If an association is found, the event is considered to originate from that PMU, or to originate closest to that linked PMU in the network. 

\section{Source Localization via FFT}
\label{sec:fft}

One of the most promising applications of PMU technology is its usage in the detection and monitoring of power system oscillations. Generally, oscillations are triggered by system faults or generator misoperation, and they can be temporary (the oscillation is damped and dies off after a couple of seconds) or sustained. Sustained oscillations, also called forced oscillations, happen when a generator (or group of generators) has imperfect or inaccurate real-time control-- it continuously injects active or reactive power that in turn excites the oscillatory modes. This type of oscillation can be stopped only when the generator is either taken offline or its controls are adjusted. 

\begin{figure}[t]
    \centering
    \includegraphics[width=0.5\columnwidth]{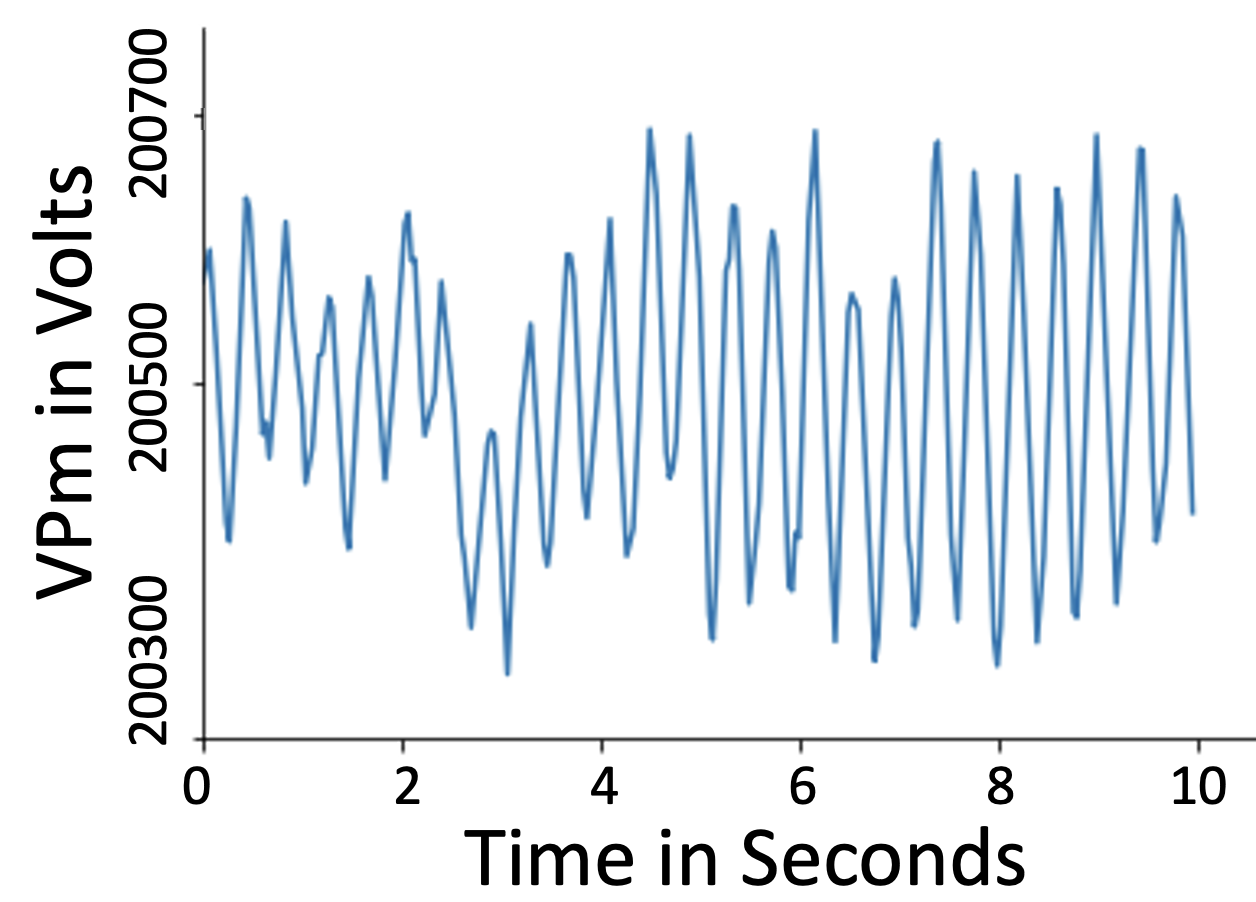}
    \caption{Oscillation of voltage magnitude (VPm) of an example PMU over a 10 second period.}
    \label{fig:osc}
\end{figure}

Oscillations can be observed from many different quantities such as voltage magnitude, frequency, and power injection (Figure~\ref{fig:osc}). Oscillations propagate throughout the system and can be observed from many different buses around the source. An important task for power system operators is to localize the oscillation source-- the bus at which the generator is exciting the system and causing the oscillatory behavior. Subsequently, corrective actions can be taken by the operators to fix the issue-- i.e., controlling the generator or taking it offline. 

\begin{figure}[t]
    \centering
    \includegraphics[width=0.6\columnwidth]{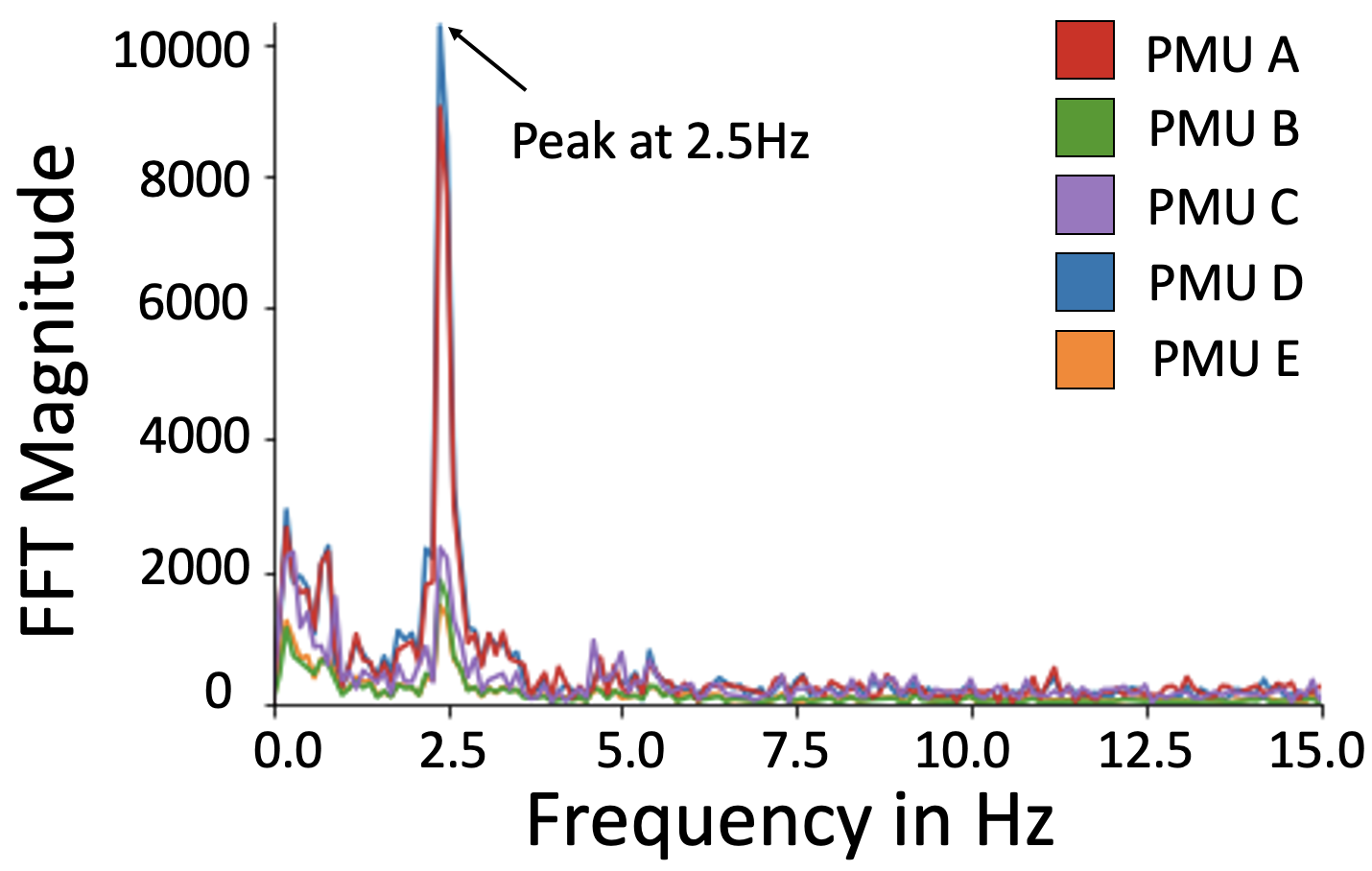}
    \caption{The FFT magnitudes for 5 PMUs, taken over a 10 second period, with voltage magnitude. The peak magnitude is seen at 2.5Hz for all PMUs, indicating a 2.5Hz oscillation event in the grid. The magnitude is highest for PMU D, indicating that it is a potential epicenter for the event.}
    \label{fig:fftosc}
\end{figure}

We, therefore, incorporate epicenter-PMU identification and juxtaposition as a key functionality of \name{}. We propose a method to localize the source of an oscillation, that fully integrates within our platform. When looking at bus voltage magnitudes during the oscillatory events in our dataset, it can be seen that the oscillations are the most prominent and clean at buses close to the source. As we get further, the voltages and the oscillations become noisier. This is due to the fact that as you get further from the source, the effect of the oscillation is damped and can be attributed to other phenomena and disturbances that are happening simultaneously in the system. Based on this observation, we perform an FFT analysis on the voltage and current signals of PMUs from different locations on the grid. By comparing the FFT magnitude at the oscillation frequency (Figure~\ref{fig:fftosc}), we can get an idea of how close each bus is to the source. We illustrate this implementation within the platform in the following sections.

\begin{figure*}[!h]
    \centering
    \includegraphics[width=0.9\textwidth]{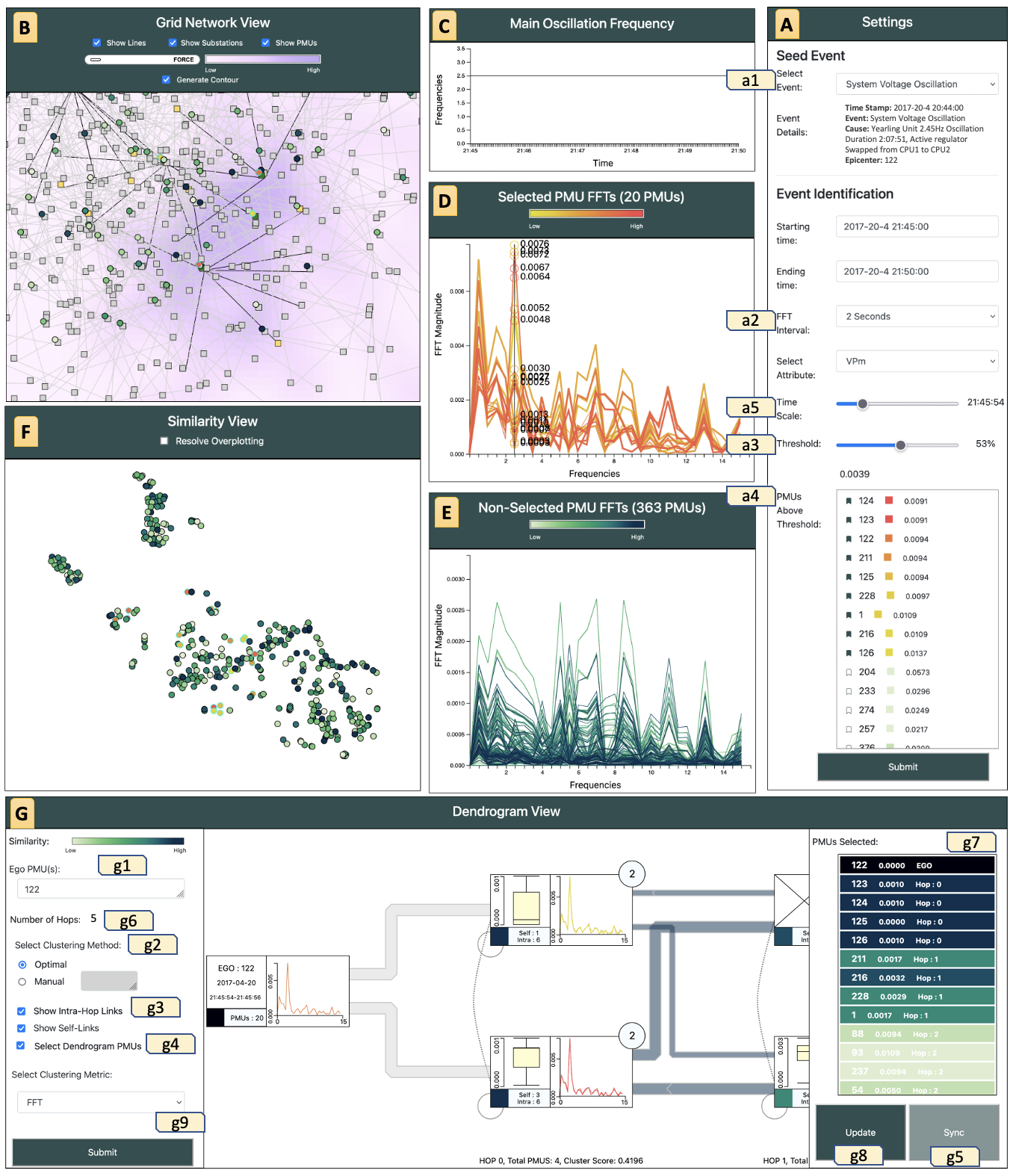}
    \caption{The PMU Tracker interface. (A) The Settings panel supports event selection and FFT analysis settings. (B) The Grid Network panel displays the power grid network with a contour layer. (C) The Main Oscillation Frequency panel displays oscillation over time-lapse. (D) The Selected PMU FFTs panel shows FFTs for PMUs under active review.
    (E) The Non-Selected PMU FFTs panel shows FFTs for remaining PMUs.
    (F) The Similarity panel shows t-SNE clustering of PMUs. (G) The dendrogram panel shows event propagation across hop-wise PMU clusters with respect to specified epicenter PMU(s). 
    }
    \label{fig:interface}
\end{figure*}

\begin{figure*}[!h]
    \centering
    \includegraphics[width=0.9\textwidth]{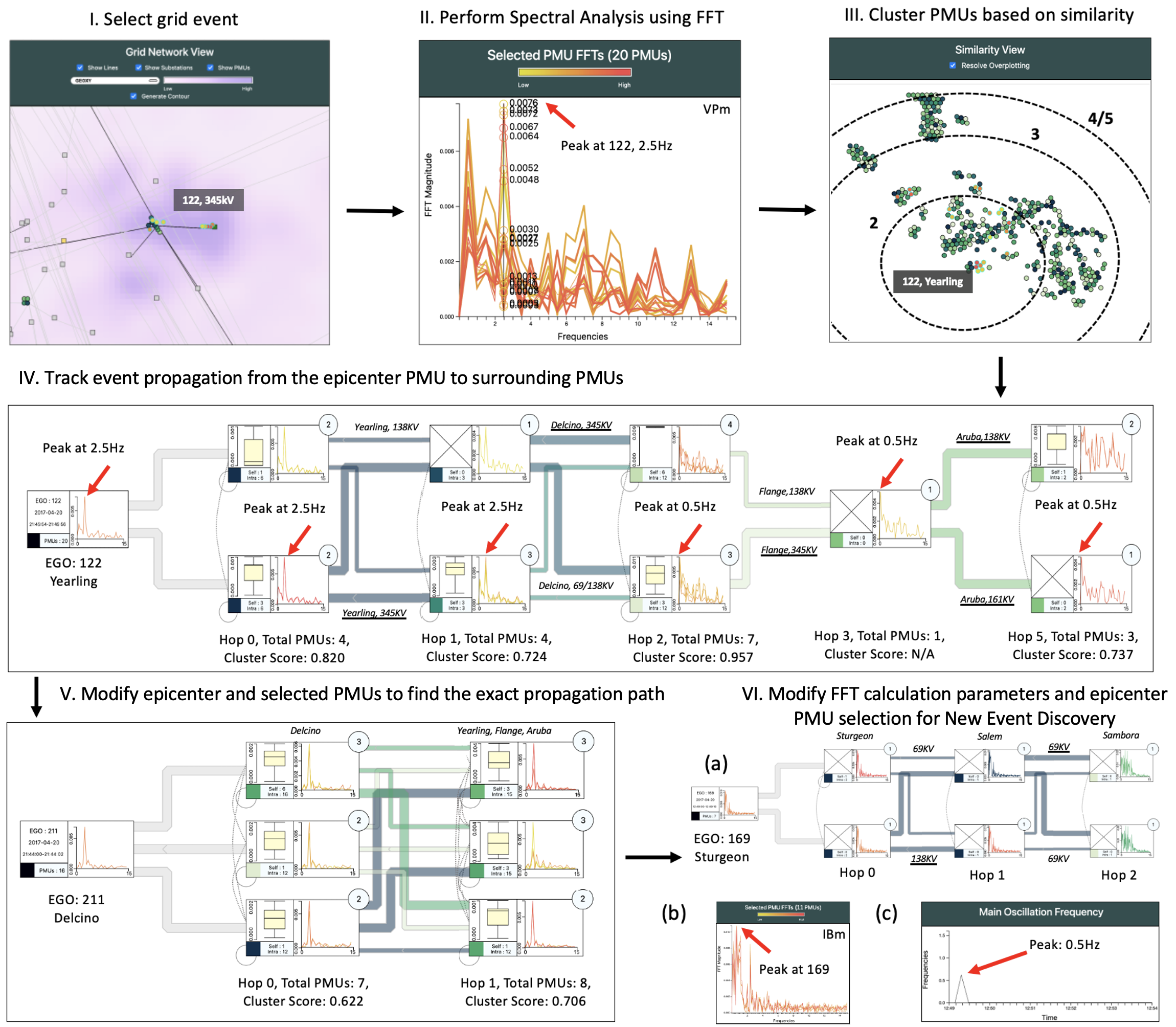}
    \caption{Usage Scenario: System Voltage Oscillation at Yearling Substation, 20:44:00, 4/20/2017. (I) Close-up of the Yearling and Delcino Substation contour, 21:45:00, 4/20/2017. (II) Spectral domain analysis is performed via FFT over voltage/current to flag PMUs of interest. (III) Similarity view without overplotting: arc annotations lay out how many hops away PMUs are, but PMUs with higher voltage are closest to the epicenter, irrespective of their cluster. (IV, V) Dendrogram technique to track event propagation from epicenter PMUs into surrounding PMUs, which are clustered based on their characteristics, as a way to understand how the event is dispersing throughout the network. (VI) Identifying a new/concurrent grid event (Transient Current Oscillation).
    }
    \label{fig:cs}
\end{figure*}

\section{Interface and Usage Scenario}
\label{sec:interface}

\name{}'s interface for event analysis and oscillation localization is shown in Figure~\ref{fig:interface}. The interface is designed to support tasks \textbf{T1}--\textbf{T4} described in Section~\ref{sec:domain_and_task_analysis} and implements the FFT approach described in Section~\ref{sec:fft}. It consists of seven linked panels (A)--(G) and supports linked highlighting: hovering on or selecting a PMU in one view highlights or selects the PMU in other views. A brief overview of the panels is as follows: (A) comprises the controls surrounding event selection and analysis parameters. (B) displays the grid network structure, as either a geographical or force-directed layout. (C) shows the main frequency of an oscillation occurring in the grid over a pre-specified time period. (D) and (E) visualize the FFT-characteristics of grid network PMUs that are either  selected or unselected (respectively) for closer inspection. (F) uses a t-SNE to cluster PMUs based on a designated epicenter PMU for the historical/identified event being analyzed. (G) houses the novel epicentric dendrogram view, which allows operators to analyze the effects of an event as it propagates outwards from a source location.

At a high level, the interface can support the following workflow: 
\begin{itemize}[noitemsep,nolistsep]
    \item Users can begin either by selecting a historical grid event for analysis or browse the dataset by specifying parameters in (A). Available parameters include the time range to query the data, FFT calculation parameters (these help define the resolution for viewing the data), settings for flagging PMUs as potentially anomalous, and options for selecting a set of PMUs.
    \item The grid network is displayed in (B)--(F), allowing users to analyze the state of the grid and compare the behaviors of flagged vs. unflagged PMUs
    \item Epicentric event analysis is supported in the dendrogram view (G). PMUs can again be selected, and these are tracked to visualize how anomalies propagate through the grid network (from a source) during a system event.
\end{itemize} 
To illustrate \name{}'s available interactions, and demonstrate how it can be used to analyze anomalous behavior, we describe the interface's features in conjunction with a usage scenario (shown in Figure~\ref{fig:cs}) analyzing a historical grid oscillation event and discovering a previously-unidentified secondary event (demarcated as \textbf{\textit{Scenario:}}~\textit{...} throughout this section's text) from the perspective of a user, \textit{Martin}, a transmission and distribution power engineer. As a note for the usage scenario, since our dataset contains historical (previously identified) events that we are analyzing, we start by selecting one in (A). However, in a real-time (or near-time) scenario (where no entry would be stored in the database), a user would begin by selecting the time frame of interest, along with a set of potentially-interesting PMUs to begin their analysis, in (A). A demo video of the interface is also available in the supplementary materials. 

\textbf{(A) Settings Panel.} 
(a1) Users first select a grid event from a dropdown list, and its details are loaded.
(a2) The timespan for FFT computation can be set along with the calculation window (2, 5, or 10 seconds), which slides across the specified start/end timestamps, and computes FFT values for the selected data attribute.
(a3) To identify anomalous PMUs, the peak FFT magnitude value over all PMUs is extracted, and a threshold percentage is specified for this peak. PMUs with values above this threshold are flagged as anomalous.
(a4) The flagged PMUs show (i) a selection icon, (ii) their ID, (iii) a color swatch showing their similarity to the epicenter PMU, and (iv) their FFT magnitude value at oscillation frequency. PMUs can be selected/de-selected from this list.
(a5) Temporal analysis (\textbf{T2}) is supported via the time-scale slider. Scrubbing this widget dynamically updates the other panels to the current time value. 

\textit{\textbf{Scenario:} }
Martin selects a System Voltage Oscillation event, with timestamp 20:44:00. The epicenter is located at PMU \#122  at the Yearling substation, and the recorded duration is about 2 hours and is described to have an oscillation frequency of 2.4Hz. He decides to initially analyze the time period of 20:44:00-20:49:00 (about one hour after the start of the event), for \textsf{VPm} (the positive sequence voltage magnitude) with a 2 second FFT window. The threshold for anomalous PMU identification is set to 100\% of the peak value, so only the epicenter PMU will be flagged. When finished, Martin clicks on the `submit' button, and panels (B)-(F) are populated.

\textbf{(B) Grid Network Panel.}
The power grid network topology is visualized using a node-link diagram. Edges denote power lines that connect buses, loads, plants, and/or transformers, which are visualized as rectangular nodes, and circular nodes denote PMUs. The network can be arranged using geographic locations, but to preserve anonymity Figure~\ref{fig:interface}(A) shows a force-directed layout.

We note that the geographical grid network is highly interconnected, with multiple PMUs often existing on the same line. Real-world usage of tools like \name{} primarily depends on this geographical view for event analysis, which follows industry standards and protocols. Event analysis involves working with a subset of PMUs of interest (that exhibit anomalous behavior) and not the juxtaposition of all network PMUs. The flagged PMUs (a4) provide context for this and are highlighted in the network view. Additionally, the white-to-purple color-map underlying the network visualizes a kernel density estimation of the peak FFT magnitude value for each PMU at the given timestep (\textbf{T4}). Such an approach follows industry standards, as a way to understand how an event's magnitude spatially distributes over the network.

We provide standard interaction techniques here, including zooming, panning, node/edge filtering, and selecting PMUs via mouse click.

\textit{\textbf{Scenario:} }
On loading the information for the selected event (Figure~\ref{fig:cs}(I) shows a zoomed-in view of the same), Martin sees in (B) that PMU \#122, the only PMU that has been selected so far, is highlighted in the network, in red. The corresponding substation, Yearling, is colored green. All lines going out of Yearling are colored black, and the connected substations are marked in yellow. The rest of the network is rendered in gray and appears de-emphasized. Martin also notes that the underlying heatmap on the view has a relatively higher (purple) intensity compared to the rest of the power grid over the region surrounding the epicenter PMU (at the Yearling substation).

\textbf{(C) Main Oscillation Frequency Panel.}
Oscillation frequency values are found by considering the position at which the maximum magnitude peak occurs at FFT calculated over every time step of the time-lapse period over all PMUs. This view allows the user to monitor the behavior of the oscillation over time: changes in the base frequency of the oscillation can indicate changes in the system response. 

\textit{
\textbf{Scenario:} }
Martin sees an initial oscillation frequency of $\sim$1 Hz shown in this view. He decides to examine a different time stamp from (A), as the event is described to have an oscillation frequency of 2.5 Hz. Martin scrubs the time slider control (a5) over the two-hour period following the event time stamp. He notes that throughout this duration in (B), the region surrounding the epicenter PMU (at the Yearling substation) has a relatively higher intensity compared to the rest of the power grid. Martin also finds that substations located 1-2 hops away from Yearling also show darker intensities compared to substations connected further away from the epicenter, and this peaks roughly an hour after the event time stamp. Martin selects 1-2 PMUs from each substation displaying these darker intensities for further review (a total of 20).

On viewing changes in (C), Martin finds that the main frequency of the oscillation gradually increases to peak at $\sim$2.5 Hz at around 21:45:00 (about one hour after the start of the event) over the VPm parameter. Following this, he sees the main frequency is maintained till 22:44:00, following which it declines to rest at $\sim$0.2 Hz over the next half hour. Based on these observations, Martin decides to, therefore, focus on the time period of 21:45:00-21:50:00 for the rest of the analysis. He resets the threshold (a4) to constitute 53\% of the peak (0.0093) at this timestamp. He can observe that 9/20 selected PMUs have been flagged as crossing this threshold.

In the case of inter-area oscillations, system events along with poorly damped electric grid facilities can lead to groups of system units such as generators, transformers, etc. oscillating simultaneously at various regions in the grid. In this situation, main frequency monitoring is important, in order to gauge the impact of an event on the system.

\textbf{(D) Selected PMU FFTs Panel.} 
Selected PMUs are plotted to show their FFT magnitudes over frequencies. The colormap assigns increasingly red shades to PMUs that have a higher correlation with the PMU that has the highest FFT magnitude, to investigate the propagation of oscillation effects.

\textit{
\textbf{Scenario:}}
Martin sees that for the new time period and PMU subset selected (as shown in Figure~\ref{fig:cs}(II)), in (D), PMU \#122 displays the highest FFT magnitude; he finds this to be consistent for the full event duration of $\sim$2 hours on scrubbing the time slider (a5). This justifies \#122's selection as the likely epicenter PMU. On juxtaposing panels (D) and (B), he also sees that (i) those PMUs topologically closer to the epicenter PMU (i.e., there are fewer hops in the network separating them) displayed higher oscillation frequency peaks over this time period, and (ii) PMUs situated at lines with the same voltage levels show similar activity.

\textbf{(E) Non-Selected PMU FFTs Panel.} 
Similarly, the FFTs of unselected PMUs are displayed in this chart. Increasingly blue shades are assigned to those PMUs that have a higher correlation with the PMU that has the highest FFT magnitude.

\textit{\textbf{Scenario:}} Martin  compares (D) and (E) to find that the highest peaks seen in (E) occur at multiple frequencies and are also of very low magnitude. He compares PMUs from (D) and (E) against flagged PMUs (a4)-- PMUs from (E) are all not flagged, and are ranked very low; he, therefore, concludes that peaks from (E) can be attributed to numerical FFT noise rather than the indication of a physical event occurring.

\textbf{(F) Similarity Panel.}
This plot uses dimensionality reduction to plot PMUs based on their similarity to each other. Specifically, t-SNE~\cite{maaten2008visualizing}, short for t-distributed stochastic neighbor embedding, is an unsupervised non-linear algorithm for embedding high-dimensional data into a low-dimensional embedding space (usually two or three dimensions). We compute the Euclidean distance between PMUs based on their FFT values and plot them using t-SNE in a two-dimensional space. PMU circles placed closer together are more similar, while PMUs farther apart are less similar. To resolve overplotting, collision detection is available, which shifts PMU points so their circles do not overlap.

\textit{\textbf{Scenario:}} Martin observes that selected PMUs are all placed in a large central cluster (Figure~\ref{fig:cs}(III)), which gives a cue to the structure of the oscillation event (i.e., it likely affects points in this cluster the most), and provides the user a pool of candidate PMUs for selection. We have annotated, via dotted concentric rings, the average hop distance from the epicenter PMU. Martin also sees that the 9 flagged PMUs are located closest to the epicenter PMU \#122 in this cluster. He notes that interestingly, some PMUs from the Flange and Delcino substations---one hop away from the Yearling substation---are not placed in this central cluster.  On examination of these PMU locations in (B) and their FFT curves (D), he observes that these PMUs do not show high FFT peaks, and are on lower voltage lines. He infers that this is likely caused by the presence of step-down transformers within these substations between buses that connect the epicenter PMU with lines going out to these external PMUs. When the voltage is stepped down, the oscillatory behavior is damped over these lower voltage lines.

\textbf{(G) Dendrogram Panel.}
The final plot is a novel dendrogram clustering panel, which uses a combination of $k$-means and hierarchical clustering to group selected PMUs into nodes, which are rendered in a dendrogram layout. 
(g1) This panel allows for any PMU (or group of PMUs) to be set as the epicenter, from which all other \textbf{selected} PMUs are juxtaposed. Each layer of the dendrogram corresponds to the hop distance from the selected epicenter PMU(s).

(g2) The clusters at each hop can be either automatically generated or manually specified (e.g., by selecting different $k$-values). A color swatch indicates the correlation between the cluster and the epicenter PMU(s), where darker indicates more correlation. Each cluster contains: (i) statistics about the number and type of links present as well as the number of cluster PMUs, (ii) the FFT line chart, and (iii) a box plot of the FFT value distribution of PMUs in the cluster. The clustering silhouette score~\cite{shahapure2020cluster}, hop number, and total PMUs per hop are also indicated at the bottom of each dendrogram layer. This view is scrollable and zoomable, which is helpful for large dendrograms with many hops. We default to the optimal clustering for each layer based on silhouette score. 

(g3) There are three types of links present in the dendrogram. (i) Self-links comprise connections between PMUs located within the same cluster, represented by a circular arc in the bottom-left corner of the node. (ii)~Intra-hop links are links present between PMUs in different clusters that belong to the same hop, represented by arced and dotted lines between clusters. (iii) Inter-hop links connect PMUs in clusters belonging to different hops, represented as colored flows. 

In most cases, the clusters between two hops will be fully connected, given that the grid network is itself densely connected. We therefore color the flows corresponding to the color swatch on the cluster of its origin; the flow thickness corresponds to the level of influence on downstream clusters, calculated based on the relative Manhattan distances of the cluster-wise averaged FFT data. To reduce clutter, self and intra-hop links can be toggled off, and the links turn at right angles.

(g4) To populate the dendrogram view, on enabling dendrogram PMU selection, the 'Sync' (g5) and 'Update' (g6) buttons are enabled. (g5) Initially, the sync button automatically includes all of the PMUs selected in (A)--(F) for clustering in the dendrogram.

\textit{\textbf{Scenario:}} Martin moves on to tracking the nature of event propagation over the selected PMUs using the dendrogram. Initially, he loads the dendrogram for the default epicenter specified, PMU \#122 (Figure~\ref{fig:cs}(IV)). He observes that as the hop distance increases, the oscillation frequency drops from 2.5Hz to 0.5Hz, with the furthest selected PMUs located 5 hops away from the epicenter. He sees that all the flagged PMUs are located a maximum of 1-2 hops away, for a threshold of 0.0093. Martin deduces that the event selected is found to mainly propagate through PMUs that are connected on high voltage lines to the Delcino substation, which is located one hop away from the epicenter substation. He bases this on the `strong' inter-hop flows between dendrogram nodes  constituting these PMUs and the epicenter cluster; these clusters are also more closely aligned (darker colored) with the epicenter cluster. Additionally, the box plots help Martin verify that clusters are formed in a meaningful manner, as they are seen to reflect uniform distributions, such that the constituent oscillating PMUs closely influence one another. 

Next, Martin updates his selection to include more PMUs located on high voltage lines at the Delcino substation, and remove unflagged PMUs. (Note: This modification still preserves the original PMU selection in previous panels.) He regenerates the dendrogram with customized clustering (3 clusters per hop). 
Martin can see in the new dendrogram (Figure~\ref{fig:cs}(V)), with PMU \#211 as the epicenter, the PMUs from the modified selection are all flagged for the set threshold. Additionally, the dominant cluster (thickest flows) in Hop 0 has PMUs located on buses for lines going from Delcino to the bus in Yearling on which the epicenter PMU is located. The FFT plots for all PMUs align and show a 2.5Hz oscillation; the box plots also show uniform distributions across all clusters. Martin can therefore verify that the system voltage oscillation propagates through PMUs on high voltage lines, over a time period of two hours.

In general, we find that 2--3 clusters per hop are generated with optimal clustering. When navigating the dendrogram view, PMUs corresponding to the view are highlighted across all other panels. In addition to event propagation tracking, the dendrogram view also enables the discovery of new grid events that might not necessarily be flagged by standard monitoring systems until they evolve to cause widespread disruptions further downstream. The flexibility afforded in potential epicenter selection drives event discovery, and constitutes the second part of our usage scenario (discovering a previously undiscovered event).

(g9) The parameters used for FFT calculation, specific to the dendrogram view, can be modified in (A). The 'Submit' button can be used to update the dendrogram according to this (panel (A)'s parameter information reverts back to the values prior to dendrogram update). This feature allows the simultaneous analysis of event characteristics recorded by PMUs at different points in time.

\textit{\textbf{Scenario:}} Martin decides to explore time periods outside the event duration, to check if any other significant anomalies or events preceded, succeed, or concurred with the oscillation event he analyzed. In this process, he discovers that at 12:49:00, on the same day as the analyzed oscillation event, a small anomaly is reflected in the IAm/IBm/ICm (positive current magnitude for phases A/B/C) of PMU \#169 at the Sturgeon  substation. Calculating FFT over 10 seconds, he finds that this is a well-damped transitory current oscillation---one that lasts for approximately 2 seconds, disappearing completely after only three cycles (as shown in Figure~\ref{fig:cs}(VI)(c)). 

Martin observes (from Figure~\ref{fig:cs}(VI)(a) and (b)) that high overlap of FFT values is seen for immediate neighbors of the epicenter PMU \#169; on further analysis, he notes that neighboring PMUs are located on low voltage lines (69kV), and are connected to step-down transformers, accounting for the highly localized oscillation (which is also symmetrical over IAm and ICm). The main frequency oscillation values for the epicenter PMU, calculated over 10 second time windows at the event time stamp, show that the system is almost completely at rest, with only a momentary sharp increase in the main frequency of the epicenter PMU to $\sim$0.7 Hz.  Martin spots that the event quickly dies out as the oscillation damps, with the system returning to an ``at rest'' state comprised of very slow oscillations that can be filtered out using a high-pass filter. He attributes these observations to the occurrence of transient events such as ``earth faults''~\cite{5744142} (comprising of line/generator tripping or high load fluctuation). These are known to give rise to current signals containing decaying DC components, and exhibiting such damped, transitory power oscillations.


\section{Evaluation via Expert Reviews}


To empirically evaluate \name{}, we 
conducted pair analytic sessions~\cite{elmqvist2015patterns} with 12 power domain experts (Table~\ref{tab:pair}); four of these practitioners regularly conduct data-driven analyses on the same regional power grid networks that constitute our dataset; for these, we used the raw (non-anonymized) interface; the other participants used the anonymized version of the interface. Each pair analytic session comprised of a visualization expert well-versed with the system functionality who “drives,” while the study participant (power domain expert) freely made analysis and investigative decisions based on their own expertise and desires. Sessions lasted between 30--45 minutes (average$=37.5$, SD$=3.26$ minutes).

When conducting a session, the visualization expert's interactions comprise largely of (i)~clarifying participants' questions regarding either the functionality of controls or the nature of data presented, (ii)~obtaining clarification from the participant regarding a specific navigation action that the participant wants to perform, if received instructions are unclear. In pair analytics, freeform verbal discussion between the driver and participant is the basis for understanding the participant’s sensemaking process as well as what specific insights are uncovered during investigation. To mitigate confounds, the driver should in no way specifically prompt or leadingly question a participant; the driver may remind the participant to verbalize their thought process periodically if the participant has stopped doing so. The visualization expert's role is therefore largely passive in the decision-making, such that maximally achievable, and objective, feedback is obtained from the participant.

We report here general feedback about the system's analytical capabilities, as well as selected quotes from experts which help illustrate general trends in feedback.

\begin{table}[!h]
    \centering
    \footnotesize
    \begin{tabularx}{\columnwidth}{XlX}
    \toprule
        \textbf{Number} & \textbf{Expert Role} & \textbf{Experience (Years)}\\    
    \midrule
         5 & Lead/Expert Engineers & 2*, 4*, 7*, 10*, 14*   \\
         3 & Post-Doctoral Researchers& 6, 6.5, 8 \\
         4 & Graduate Students &  1.5*, 2, 2.5, 3.5*\\
    \midrule
    \end{tabularx}
    \caption{Demographics for participants of pair analytic sessions. All experts surveyed worked in either systems protection and control or contingency analysis for electrical utilities, as either current*/former Power System Engineers.}
    \label{tab:pair}
\end{table}

Feedback was generally positive, with all experts particularly excited about the dendrogram panel as a new way to characterize the impact and evolution of oscillations on both an event's epicenter PMU and its surrounding PMUs. ``\textit{ [The dendrogram] in simulation-based event analysis since the flows make it easy to follow the relations between clusters.}'' The epicentric nature of the analysis was also appreciated: ``\textit{[using] a combination of epicenters and changing the time scale will help with concurrent event identification as well,}'' and ``\textit{coarse-grained mapping to geographical regions of interest where multiple PMUs seem to move together is afforded.}''

One engineer remarked that, when used in association with the network graph contour and similarity panel, the dendrogram ``\textit{allows for easy extensibility to other spectral analysis techniques like modal analysis for multiple events,}'' and that ``\textit{event classification can be implemented based on oscillation modes, with the isolation of the most affected PMUs.}'' Another researcher suggested that, ``\textit{line flows and line parameter estimates can be simulated on the network and dendrogram panels to map out control protocols.}'' These comments help demonstrate that, not only was the interface seen as useful, but it opened up new possibilities for visualizing and analyzing grid data.

The mapping of spatial PMU data with time-varying attribute values was found to be well-implemented. ``\textit{It's difficult usually, going between the map and everything else. Here the color-coding of both points [PMUs in FFT panels] and the clusters formed for the last [dendrogram] helps you hop around the network.}'' Some experts remarked that adding alternative or automatic selection interactions might minimize operator overhead further.``\textit{[network] PMUs could be lassoed, with a limit on the number of PMUs per substation, so we could just drag a net on the whole area. }''``\textit{[Flagging] is a neat idea to tie in with protocols in place-- we get flashing notifications on screen. Maybe extending that to auto-prune everything that doesn't matter would save time if it's a high-stakes situation. High-quality communication, yes, but also speed in relaying to downstream operators to shift or cut transmission to protect equipment.}''

One significant concern posed by a couple of experts concerned the analysis of concurrent events. ``[We can] \textit{pick out anomalies and not the lower oscillation frequency noises that we might see, so we can separate out an event, fine. What do I do if I have overlapping ones? [The big drops] happen when there are multiple failure points. So we might have multiple chokes, which could stem from multiple epicenters. Not saying that we can't find them all, just speed might be a problem since we're doing one thing at a time. Unless we have multiple screens with different things being concurrently tested on the exact same pipeline.}'' ``\textit{When we have [concurrent] events like that, would we see a wraparound in the dendrogram at some point? We could, especially when you start setting multiple epicenter PMUs in your initial epi-cluster. So a gradation of flows, or some sort of demarcation based on which epicenter it's closest to would help us with stemming that effect.}'' We discuss this limitation in Section~\ref{sec:limitations}.

However, many experts also felt that the adaptability of the dendrogram panel to use multiple PMUs as epicenters might help address concurrent event characteristics. ``\textit{I like that you can put any PMUs you want into the epicenter cluster. If I put unrelated ones in, I should know there's another anomaly somewhere if the dendrogram flow strengths look off. So I'd have to individually look at flows for each PMU I just grouped in, one or two hops away. We don't usually see very widespread repercussions from a single fault, it has to be a lot working together for one thing to pop up 10 links away on the network.}''\textit{The functionality is very cool [for the dendrogram], particularly customizing clusters. It just needs a couple of tweaks to show different gradations in flow if you have multiple epicenters simultaneously. Maybe a toggle saying flows for the first PMU, for the second, for both, [you know] something easy to switch between a set of scenarios for potential epicenters. That would be good for discovery. I could open up multiple dendrograms below each other to compare for simultaneous faults.}``

Experts also suggested integrating real-time processing functionalities in the future: ``\textit{It's pretty neat, with the flagging and the tree taking you through the network in a more intelligent manner to not overwhelm you. I think putting this with a streaming source would be great you could see everything move and freeze when you needed a closer look at something flagged.}'' The scalability of PMU Tracker was also found to support the streamlining of protocols followed in event analysis, though experts made a few suggestions to extend it further. ``\textit{We [in most cases] only concurrently monitor and analyze 1-2 PMUs from anywhere between 5-10 substations across the grid to track an event’s propagation. [PMU Tracker] can easily support that, I've got 32 [on screen] here.}"``\textit{I think the busiest panels are the FFTs and the dendrogram. The FFTs sort themselves out mostly, I only need to look at the x-axis position and the highest peak so I don't mind the clutter... [additionally] I have the ranked and flagged PMUs in the settings and the similarity clusters.}"``\textit{It would be good to vary say the opacity or something or group the lines into clusters sort of like what you've done in the dendrogram... [take the] average to scale up more.}"

\section{Discussion}
\label{sec:disc}

Based on our case study and expert reviews, we believe that \name{} (i) viably supports tasks \textbf{T1}--\textbf{T4}, and (ii) handles most of the domain challenges outlined in Table~\ref{tab:survey_challenges}. Here, we discuss some key takeaways and lessons learned from our design and evaluation process and discuss current limitations and future directions for tools like \name{}.

\subsection{Implications for Power Systems and Other Domains}

\textbf{Event Tracking is Faster with a Tailored View.}
The experts agreed that the dendrogram panel streamlined PMU selection and analysis for event identification and epicentric propagation tracing, particularly when compared to non-epicentric monitoring interfaces~\cite{6863485}, which mimic topological data display for the power grid's network of PMUs.  In part, this reflects how custom-designed visualizations can better support data-driven power grid analytic needs, instead of relying on ``traditional'' techniques, such as the node-link diagrams (which are commonly used in control rooms) that provide top-down perspectives of the grid.

\textbf{Support for High-Level Insights.}

An important consideration in the design and development of \name{} is the intended user population. Power grid engineers and operators in our expert reviews were generally familiar with the ``standard'' visualization techniques (node-link diagrams, line charts, etc.) employed in the interface, but the epicentric cluster dendrogram view, as a novel visualization technique, was unfamiliar to them. The use of other familiar visualizations (along with linked highlighting) helped them to contextualize how to use the dendrogram panel, reducing the learning curve associated with it, and improved their ability to explore PMUs and derive insights.

The inclusion of multiple PMUs in the epicenter cluster to generate the dendrogram supports multiple perspectives of the grid network for analysis. By observing the directionality of flows and customizing cluster distributions, users could perform association and comparison tasks between PMUs during event identification and tracing. Some of our experts recommended adding features such as a toggle or opening multiple views to contrast differently centered dendrograms against each other to promote easy concurrent event discovery. However, the panel was largely appreciated for the functionality it affords in tracing event propagation, both for historical events in the dataset and when identifying previously-unknown incidents.

\textbf{Generalizability.}
We have begun testing our system on additional time-varying topological domains, such as seismic sensor fields~\cite{fauvel2019earthquake}, to help understand its generalizability for epicentric analysis tasks. Though seismograms do not form a physical network, the magnitude and propagation of tremors felt across the various sensors acts in an epicentric manner (flowing outward from the epicenter of the occurring earthquake). While more extensive evaluation is required to reach concrete estimations of the interface's efficacy in other types of data domains, our initial investigations are in line with those of power data, indicating that the interface (and in particular, the dendrogram panel) are likely adept at analyzing other time-varying sensor network datasets. As the other techniques in \name{} (node-link diagram, t-SNE, FFT plots) have been used in other tools for analyzing these types of data, it indicates that \name{} is applicable not just for power grid data, but for other relevant networks with similar epicentric event and data considerations.

\subsection{Current System Limitations and Future Directions} 
\label{sec:limitations}

We are currently working with industry collaborators to expand the capabilities of \name{}, notably for \textit{streaming event analysis} and even \textit{prospective event analysis} (predicting and assessing events that have not yet occurred). This is because a primary current limitation in \name{} is that it does not handle \textit{streaming PMU data for real-time decision-making}, which is a significant and acknowledged industry problem (e.g., it was noted during expert review sessions). Transitioning \name{} to support real-time event analysis is a non-trivial software engineering and research challenge, which will require both algorithmic integration, such as a modal or spectral analysis that can flag anomalous behavior in real-time, and backend support for scalable, on-the-fly data processing, via databases such as BTrDB~\cite{andersen2016btrdb}. 

A significant concern in real-time event scenarios is that the epicenter PMU is not known \textit{a priori}, and recent research has shown that automated anomaly approaches to identify epicenter PMUs (e.g.,\cite{8082533,8353135,4275462}) are ripe for misidentification~\cite{9043670}. Human-in-the-loop visual interfaces provide a viable strategy in this scenario, as users can employ domain expertise to review algorithm recommendations when determining the source of an event.

Another concern raised by experts during the pair analytic sessions was the difficulty of using the current system for concurrent event analysis. However, as mentioned by the experts themselves, features such as opening multiple views or toggles for multi-PMU clusters to change flow visibility against single epicenters could support such analysis. We believe that while our system can scale beyond 30-40 PMUs, future work can investigate ways to visualize many dozens or even hundreds of PMUs for large-scale grid events that could concurrently occur, which would potentially overload some of the views in PMU Tracker. Additionally, PMU Tracker can be modified to explicitly notify users and auto-select PMUs of priority (in the case of streaming data) for monitoring, to further reduce operator overhead. 

We also plan to explore integrating additional data streams as a way to augment the analysis of grid events, such as weather data, electrical market data, and social media. While omitting these data types likely limits the holistic analysis that can be performed, such data is difficult to integrate and synthesize in a way that effectively supports real-time usage scenarios. Despite this, such data streams may provide beneficial information, especially when dealing with events as they happen. For example, on December 27th, 2018, an explosion damaged a transformer in an electrical substation in New York, leaving over 10,000 customers without power~\cite{astoria_story}. Integrating social media data streams could help operators quickly assess and validate the extent and impact of such events.

Finally, one aspect of event analysis that currently is currently not supported by PMU Tracker is recommendations that can be made to operators in order to resolve an event. Our goal with this system is to streamline the event discovery and tracking phases of event analysis, in order to successfully understand an event and estimate the extent of disruption it can potentially cause. PMU Tracker is intended to serve as an intermediate link between existing industry platforms dedicated to wide-area monitoring systems and implementation of event resolution protocols, with a focus on streamlining the process of event localization.

\section{Conclusion}

We present \name{}, a software tool for event localization and tracing based on power system visualization and the analysis of PMU-equipped smart grid data. We identify a set of tasks to support event analysis and design a coordinated, human-in-the-loop visualization interface to support these tasks. In particular, we utilize signal processing to localize the source of oscillation events, paired with a novel epicentric cluster dendrogram visualization to understand how the events propagate through the power grid network. We validate \name{}'s design and discuss how it can be generalized for epicentric analysis in other time-varying sensor networks (such as seismogram data).

\acknowledgments{
This research was supported by the U.S. National Science Foundation through grant OAC-1934766.
}

\bibliographystyle{abbrv-doi}

\bibliography{template}
\end{document}